\documentclass[preprint,12pt]{aastex}

\usepackage{epsf}
\usepackage{landscape}
\def\et{et al.}
\def\ab{$\sim$}

\def\K{K}

\def\hu{km~s$^{-1}$~Mpc$^{-1}$}

\begin{document}
\slugcomment{Final version for ApJ press, 27 Sep 2001}
\vskip 0.5in
\title{Are Starburst Galaxies the Hosts of Gamma-Ray Bursts ?\altaffilmark{1}}
\author{R. Chary\altaffilmark{2,4}, E. E. Becklin\altaffilmark{2},
L. Armus\altaffilmark{3}}
\altaffiltext{1}{Based on observations at the W. M. Keck Observatory}
\altaffiltext{2}{Division of Astronomy \& Astrophysics, 
University of California, Los Angeles, CA 90095-1562}
\altaffiltext{3}{SIRTF Science Center, California Institute of Technology, Pasadena, CA 91125}
\altaffiltext{4}{Present Address: Department of Astronomy \& Astrophysics, University of
California, Santa Cruz, CA 95064; email:rchary@ucolick.org}

\vskip 2in
\begin{abstract}

We present deep 2.2~$\mu m$ imaging of twelve gamma-ray burst host galaxies. 
Template spectral energy distributions are fit to the multiband photometry between
visible and near-infrared wavelengths
to derive a better constraint on the stellar mass of these galaxies.
The internal extinction in the host galaxies is estimated using the rest-frame
ultraviolet (UV) slope. We find that the extinction corrected star-formation rates (SFRs) of the galaxies
are significantly larger than rates derived 
from emission lines in the visible or the UV continuum.
The ratio between the extinction corrected SFRs and
stellar mass for 7 of the host galaxies is high compared to local starbursts
and 3 of the hosts have derived far-infrared luminosities comparable to infrared luminous galaxies.
In addition, 
existing observational data reveal that at least 6 of the 11 putative hosts seem to be
disturbed or have companion galaxies within
a projected angular separation of $\sim$2.5$\arcsec$. 
If we assume that the host and the companion are at similar redshifts,
this corresponds to a physical separation of less than 20~kpc, 
providing some evidence for an ongoing/recent tidal encounter. 
We conclude that tidally-induced starbursts such as those found in infrared
luminous galaxies might be popular birthplaces for gamma-ray bursts.
The age of the stellar population in 4 out of 6 galaxies is rather young, 
of order 10~Myr. This favors models where gamma-ray bursts 
are due to the core-collapse of isolated, massive stars and explosion
of the resultant black hole-accretion disk system. 

\end{abstract}


\clearpage

\section{Introduction}

The detection of decaying X-ray, visible, near-infrared and radio transients 
associated with long duration ($>$ 2~sec) gamma ray bursts (GRBs) has resulted in an accurate
localization of the burst positions in the sky \citep[e.g. ][]{kul00, piro00, fra97, vanP97}. 
Redshift studies of the transients have confirmed the cosmological 
origin of at least some
of the bursts.  After the transients have faded,
it has been possible to search for an underlying galaxy which may
be associated with the bursts as predicted
in most cosmological scenarios for GRBs.
The searches have been successful at visible wavelengths with an
underlying galaxy (hereafter, a `host') 
being detected for most of the bursts which have associated
transients at other wavelengths \citep[e.g.][]{fra99, blo99a}. 
However, this has provided limited insight into the progenitors
of the bursts although observations of the transients have provided many
critical constraints for theoretical models of GRBs especially with respect to
the energetics and radiation mechanisms in an expanding fireball
\citep[e.g.][]{mes97}.
Since it seems impossible to directly observe the progenitors of the bursts,
studying the characteristics of the underlying host galaxy
may provide some insight into their origin. For example, if the progenitors
of GRBs were massive stars that collapse into black holes \citep{woo93},
GRBs would tend to lie in sites of active star-formation. On the other hand, models
that involve coalescing neutron stars or black holes \citep{nar92} would tend to
produce a significant number of bursts that are offset from the nucleus of their hosts
because of the large kick imparted to the remnant from its supernova. A
third likely possibility is that if
GRBs originate in supermassive black holes that
constitute the central engine of quasars, one would 
find the burst
positions to be located at the nucleus of galaxies \citep{rol94}.

\citet{blo01} have 
compared the observed offset of GRBs from the centers of their host galaxy 
to the predicted offset for two different GRB models. They find that the median 
observed offset for the GRBs has a projected distance of 3.1 kpc and that the 
distribution is inconsistent with a GRB model involving delayed coalescence of stellar remnants
which would result in significantly larger offsets between the bursts and the host galaxies \citep{blo99}.
However, the observed offsets are consistent with a ``collapsar'' model involving the explosion
of a massive star through a black hole-accretion disk system
\citep{mac99, fry99} provided one assumes that massive star formation takes
place in an exponential disk. Further tentative evidence for a stellar origin
for the bursts comes from the apparent connection between 
SN1998bw and GRB980425 \citep{gal99} as well as
analysis of the light curve at visible bands of GRB980326 and GRB970228 \citep{blo99b, gal00}
both of which showed an increase in brightness \ab3 weeks after the burst. The time evolution
of the optical transient brightness for these two cases
was interpreted as a power-law GRB light curve coadded on
a Type Ic supernova light curve at the appropriate redshift.
The two strongest pieces of evidence though, for an association 
between GRBs and massive
stars comes from the detection of the host galaxy
of GRB980703 at radio wavelengths \citep{berg01} and 
the X-ray detection of 
an iron line and the iron-recombination edge in GRB991216 \citep{piro00b}.
The radio detection of the GRB980703 host implies that the galaxy is undergoing a
violent starburst with a star-formation rate in massive stars of $\sim$90~M$_\sun$/yr.
The detection of the iron line suggests that the GRB progenitor exploded in a mass-rich
environment which probably resulted from recent outflows from the progenitor star.
This appears to disfavor
merging black hole-neutron star systems as the cause for GRBs
since the stellar progenitors of these 
objects would have moved away from their original environments over the merger timescale.

Recent analysis of the cosmic infrared background and mid- and far-infrared galaxy
number counts have revealed that the bulk of the high-redshift star formation takes place in dust
enshrouded regions \citep{cha01, xu01} and that the UV determined
SFRs are a factor of 3$-$7 lower than the far-infrared determined SFRs.
Furthermore, at least 70\% of this star formation takes place in infrared luminous
galaxies with 
${\rm L_{IR}}={\rm L}(8-1000~\mu {\rm m}) \ge 10^{11} {\rm L}_{\sun}$ and which have $\sim$90\% of their
bolometric luminosity being emitted at far-infrared ($40-500~\mu {\rm m}$) wavelengths.
These objects are inconspicuous at visible wavelengths except for the fact that
a large fraction of them show irregular morphologies and strong evidence for interactions.
Thus, if GRBs are associated with exploding massive stars, one would expect
their host galaxies to be primarily infrared luminous galaxies. 

Interestingly, except for the hosts of GRB980703 and GRB980613 \citep{berg01, sgd00}, none
 of the other host galaxies studied so far show evidence of starbursts or
abnormally high star formation rates based on measurements of the UV continuum or \ion{O}{2}
line strength
(e.g. Bloom et al. 1999a, Odewahn et al. 1998). Furthermore, none of the hosts have been detected
by SCUBA \citep{Smi99}, the submillimeter instrument on the JCMT which has been extremely
successful in detecting high-redshift ultraluminous infrared galaxies (ULIGs) \footnote{
ULIGs have ${\rm L_{IR}>10^{12}~L_{\sun}}$ while luminous infrared galaxies (LIGs) have
$10^{12}>{\rm L_{IR}>10^{11}~L_{\sun}}$.}.
However, the SCUBA observations were typically sensitive to an 850~$\mu m$ flux density
of 3$\sigma$\ab6~mJy restricting their ability to only detect galaxies at the tip
of the far-infrared luminosity function which account for less than 10\% of the
dust enshrouded star-formation.

Near-infrared ($1.0<\lambda<2.5~\mu {\rm m}$)
observations are well suited for establishing the characteristics
of the hosts.  At $z<$2, K-band ($\lambda=2.2~\mu {\rm m}$) observations 
trace the rest-frame stellar
mass of galaxies which would provide a better estimate of galaxy morphology.
At higher redshifts, the peak of the blue light from recent star-formation
would be redshifted into the K-band providing a better estimate of 
the unobscured star-formation
rate.  This, in conjunction with the visible light observations, which trace the
rest-frame UV light
provides an estimate of the rest-frame UV$-$visible slope and
thereby the amount of dust obscuration in the galaxy.
Lastly, for low galactic latitude fields, near-infrared
observations are less affected by Galactic extinction which 
is patchy and whose distribution is not well understood.

Thus to provide a better study of GRB host galaxy properties,
we obtained deep \K-band 
images of some GRB fields whose positions were accurately determined
by the detection of transients at other wavelengths.
These are GRB970228, GRB970508, GRB971214, GRB980326,
GRB980329, GRB980519, GRB980613, GRB980703, GRB981220, GRB990123,
GRB991208 and GRB000301C.
The transient had faded to below detection limits by the time of the observations
so the photometry 
is uncontaminated by the light from the afterglow. Most of these sources
have also been shown to have an underlying host galaxy at 
visible wavelengths. 
We use our observations in conjunction with the existing photometry and
redshift data in the literature to constrain 
star formation rates
and derive information about the characteristics of GRB
hosts.

\section{Observations and Reduction}

The near-infrared observations were made using the NIRC instrument 
(Matthews \& Soifer 1994) at the f/25 focus
of the 10m Keck I telescope on UT Nov 29-Dec 1 1998, Jan 11-12 1999, 
Jan 29-Jan 30 1999, Apr 28-29 1999 and Apr 19 2000. 
NIRC contains a 256$\times$256 InSb array with
a pixel size of 0.15$\arcsec$ implying a field of view of 
38.4$\arcsec\times38.4\arcsec$. A standard K filter
($\lambda_{0}=2.21\micron$, $\Delta\lambda$=0.4$\micron$) 
and K$_{\rm s}$~filter ($\lambda$=2.16$\micron$, $\Delta\lambda$=0.3$\micron$)
was used.  The nights of Nov-Dec 1998 and Apr 28, 1999 were not 
photometric. Observations
were made by dithering in a random pattern with integrations of
duration 8$\times$15s (coadds$\times$exposure) or
12$\times$5s per position. 
Table 1 summarizes the observations, along with typical seeing
values and sensitivities. 
Seeing values were determined from the full width at half maximum (FWHM) of
field star profiles. Several standards from the {\it Hubble Space 
Telescope} faint standard lists were observed over each night \citep{per98}.

Clipped averages of the dark current images were subtracted from
all the frames taken over a night. Bad pixels were masked
as were bright field sources.
These dark-subtracted masked frames
were then median-ed together and normalized to create a sky superflat.
In the reduction of the individual object frames, appropriate sky frames
were generated from the masked images to minimize any pattern
that may persist in the reduced frames
because of the presence of bright sources in the field. The
generated sky was normalized to the object frame by the ratio of the modes
and subtracted. 
A second order sky was also fit to the individual frames and subtracted.
The dark and sky subtracted, flatfielded images were
stacked and averaged by aligning on one of the 
bright, unsaturated, field stars. 
The typical point source sensitivity in our final, reduced images
was 1$\sigma$$\sim$24 mag in the K-band. 

In almost all the cases, deep visible light observations have indicated the
presence of an underlying host galaxy (Table 2). 
Photometry on our infrared images was performed in a beam centered on the position
of the host galaxy. If 
the host was undetected in the visible light images, the beam was centered
at the position of the radio or visible light transient which was
determined by aligning an earlier epoch observation which detected the transient
and other reference objects in the field with our final reduced frames.
Corrections for the finite beamsize was applied by curve of growth analysis
of one of the field star profiles.
Atmospheric extinction corrections ($\sim$0.05~mags/airmass)
were also applied.
The few observations that were done in non-photometric conditions had other photometric data
available (except GRB980329) which served as a basis for performing relative photometry with respect to
field objects. For GRB980329, the photometry was performed using the average zero point from
two observations of a standard which were taken just prior and 2 hours prior to the target frames.
Any variation of the zero point during the observations of the burst field
was monitored by photometry on a field star. The
zero point varied by less than 0.1~mag over the period of the observations and hence we conclude
that our derived upper limit is reliable.

Table 2 shows the observed parameters of GRB host galaxies\footnote{It
should be noted that some of the magnitudes are based on observations
with the HST/STIS
clear filter and the transformation to standard V-magnitudes is
a function of the color of the objects and therefore quite uncertain.}. 
Wherever possible, observations that are 
further in time from the burst i.e. ``late-time'', are listed to minimize
the effect of afterglow contamination. In cases where late-time visible
photometry is not available, the published magnitudes
for the host
have been derived from a fit to the integrated light of the transient and the host,
assuming a power-law decay of the transient and a constant brightness for the
underlying host. 
For the analysis,
the magnitudes were corrected for Galactic extinction based on the model of
Schlegel et al. (1998). 
The Galactic extinction values are shown in Table 3 and an R$_{\rm V}$=A$_{\rm V}$/E(B$-$V) of 3.1 was
adopted. Extinction at other wavelengths was calculated using the Galactic extinction
curve of Mathis (1990). The derived luminosity of the galaxies between
UV and near-infrared wavelengths that is shown in Table 3, is
derived by integrating over the spectral energy distribution derived from the multiband photometry.
For the luminosity distances, an $\Omega_{\rm m,0}$=0.3, $\Lambda_{0}$=0.7,
H$_{0}$=75~\hu cosmology was adopted.

\section{Characteristics of GRB Hosts}

\subsection{Morphology and Companions}

The effect of redshift induced band-shifting on morphology of galaxies has been well
demonstrated in the Hubble Deep Field-North \citep{bun00, med01, corb01}. High redshift galaxies appear
to be more irregular at observed blue wavelengths than 
at red/near-infrared
wavelengths. This is because
$z>1$ galaxies have their rest-frame UV light, which is dominated
by patchy star-formation, redshifted into the visible bandpasses. In addition,
extinction in the UV is much stronger than in the visible, amplifying the irregular
distribution of UV light. In contrast, 
the rest-frame visible and near-infrared light from these galaxies
which is dominated by main sequence stars
and K giants respectively, is less extincted and is redshifted into the K-band.
Of the GRB hosts listed in Table 2, GRB970228, GRB971214, 
GRB980519, GRB980613 and GRB990123 show somewhat distorted morphologies at
visible wavelengths (see references in Table 2). In comparison, GRB970228, GRB980519, GRB980613, 
GRB981220, GRB990123
are significantly extended in the K-band images. In addition, 8 of the 12 bursts
(GRB970228, GRB971214, GRB980519, GRB980613, GRB980703, GRB981220, GRB990123, GRB991208, GRB000301C)
show the presence of one or more companion galaxies within \ab$2.5\arcsec$ of the burst position (Figure 1).
At a depth of K=23~mag, the number density of galaxies is
about 10$^{5}$ deg$^{-2}$ or 7.7$\times 10^{-3}$ arcsec$^{-2}$ \citep{sgd95}.
Hence, the probability that
7 of the reliably determined hosts (except GRB981220)
would have companion galaxies within a 2.5$\arcsec$
radius represents an overdensity by an order of magnitude and seems intriguing.
In comparison, only about 20\% of the field galaxies observed at visible wavelengths seem to show
evidence of a companion within the same projected distance at redshifts $\sim$1 \citep{Lef00}.
Although the sample size is small,
at least for some of the listed cases, it seems reasonable to surmise
that the companion galaxy is tidally interacting
with the host, inducing
a starburst that could provide the progenitors of gamma-ray bursts.

\subsection{Extinction-corrected Star Formation Rates}

We now attempt to derive physical properties of the host galaxies such as the
amount of internal extinction and the extinction-corrected SFR
from their rest-frame UV photometry using the $\beta-$slope technique described in \citet{Meu99}.

The $\beta-$slope technique was developed because a large fraction of the visible/UV light in 
starbursts/infrared
luminous galaxies is thermally reprocessed by dust into the far-infrared.
The relative attenuation of the visible/UV light is dependent on the relative
extinction properties of dust at those wavelengths. So, a measurement of the UV slope 
traces the amount of dust extinction in the galaxy and except for the most luminous
far-infrared sources, provides a good measure of the star-formation obscured by dust.
The technique fails for ULIGs in that it provides only a strong lower limit to the opacity
and thereby the amount of 
dust-obscured star-formation. This is because ULIGs have regions of
very high dust opacity ($\tau_{\rm UV}\gg1$) where all the 
UV photons are thermally reprocessed by the
dust grains. As a result, the UV-slope of a ULIG is sensitive only to the light coming from
optically thin regions ($\tau_{\rm UV}\lesssim$ 1) and does not trace the rate of star-formation
in the high opacity regions.

Using the photometry in Table 2, we determined 
the rest-frame UV slope ($\beta$) shortward of about 
350~nm which for most
galaxies, since 
they are at $z>1$, is derived from a power law fit to the B, V and R-band observed
magnitudes. It should be emphasized that this is an extremely liberal definition of
the UV-slope primarily because the broadband photometry could be contaminated by absorption
and emission lines. Secondly, the uncertainty in the photometry is substantial and 
the measurements are consistent with a wide range of UV slopes which translate to large
uncertainties in the derived extinction values.
For example, a 20\% error in the UV-slope measurement results in a 0.4~mag error in the
derived extinction at UV wavelengths and a 50\% uncertainty in the 
dust-obscured SFR value. This effect is particularly important for GRB970228, GRB971214
and GRB991208, all of which have only observations at two filters tracing the UV light.
Lastly, for the low redshift objects, the V-band traces rest-frame 320~nm which
is much longer than the 250~nm adopted by \citet{Meu99} for performing the calibration
between the UV-slope and the FIR luminosity. This does not seem to be a significant problem
if the starburst extinction law of \citet{cal00} is adopted. This is because, for a young
stellar population, even
large values of extinction such as A$_{{\rm V}}\sim2$~mag, result in an observed spectrum which can
be well fit by a single power law between 100~nm and 400~nm. 

After estimating the UV-slope, we use the relationship
derived by \citet{Meu99}:
\begin{eqnarray}
{\rm A}_{1600} = 4.43 + 1.99\beta\\
\log [{\rm L_{FIR}}/\nu {\rm L}_{\nu}(1600)] = \log (10^{0.4{\rm A_{1600}}}-1)+0.076
\end{eqnarray}
to determine the far-infrared luminosity of the galaxy. 
In the above equations, A$_{1600}$ is the extinction at 160~nm which in the \citet{cal00}
extinction law is \ab2.4~A$_{\rm V}$ and in the Galactic extinction law is \ab2.6~A$_{\rm V}$,
where A$_{\rm V}$ is the extinction in the V-band.
The far-infrared luminosity
is typically about 83\% of the total infrared luminosity which can then be transformed
into a SFR. The calibration of \citet{Kenn98} yields:
\begin{equation}
\rho~({\rm M_{\sun}~yr^{-1}}) = 1.71\times10^{-10}~({\rm L_{IR}}/{\rm L}_{\sun})
\end{equation}
From this, we can derive a lower limit to the true SFR which is
typically higher than the UV-derived value. The value is a lower limit because,
for reasons mentioned earlier,
\citet{Meu00} find that the $\beta-$slope technique underestimates
the far-infrared
luminosity of ULIGs which contribute $\sim$30\% of the global star-formation at high redshift.
Thus, we derive dust-enshrouded SFRs of 1.6, 2.5, 70, 30, 7, 8 M$_{\sun}$/yr for 
the host galaxies of
GRB970228, GRB970508, GRB980613, GRB980703, GRB990123, GRB991208. 
The corresponding A$_{1600}$ values are 2.3, 2.5, 4.5, 2.2, 1.2, and 3.8 mag while the
predicted lower limits to their infrared luminosity
L$_{\rm IR}$ are $9.3\times10^{9}$, $1.4\times10^{10}$, $4\times10^{11}$, $1.7\times10^{11}$, $3.8\times10^{10}$ and
$4.5\times10^{10}$~L$_{\sun}$ respectively.
The object
with the one of the highest SFR here, GRB980703, was also detected in the radio by
\citet{berg01} but the SFR in massive stars derived from the radio luminosity is about 
a factor of 3~higher.

There is some uncertainty in the V-band photometry of GRB971214 which
differs by about a magnitude between \citet{ode98} and \citet{sok01}. If 
we adopt a geometric mean of the two values which corresponds to V$=26.2\pm0.3$~mag, then 
the rest-frame UV slope suggests
an average A$_{\rm V}\sim1.8$~mag. 
An empirically derived correction 
of $\beta_{\rm phot}-\beta_{\rm spec}=0.5$ has been applied in the 
derivation of the UV-slope of GRB971214 to account
for the difference between a spectroscopically
derived UV-slope and the slope derived from broadband photometry \citep{Meu99}. 
This difference is due to stellar and interstellar 
absorption features which redden the flux at these UV wavelengths. 
The derived infrared luminosity corresponding to this UV-slope is $2\times10^{12}$~L$_{\sun}$
and the SFR is 340~M$_{\sun}$/yr
which would make it a starburst comparable to the host of GRB980703. 
Further high quality multiband photometry at visible wavelengths is required to assess the
accuracy of this unusually high number since our value is a lower limit as explained earlier, 
and is barely consistent with the submillimeter
upper limit of \citet{Smi99}. The possibility of AGN contamination
also cannot be ruled out although the spectra
presented in \citet{kul98} do not show any broad lines. 

Thus, we find that the internal extinction in GRB hosts
calculated from their UV slope
is significant, with an average value of A$_{\rm V}$\ab1.2~mag which is similar to that
derived by \citet{sok01}.
As a result, the dust-enshrouded SFR values derived above are typically higher than those
derived from observations of the UV continuum (Table 3). However, the total (obscured+UV)
star-formation rate in these galaxies is still not unusually high.
It is useful to note that although only 2 of the galaxies have derived
L$_{{\rm IR}}>10^{11}$~L$_{\sun}$ and high resultant SFRs, the 
ratio between the dust-obscured SFR and the unobscured
SFR (i.e. the ratio between column 7 and column 6 in Table 3, excluding GRB971214) for 
the GRB hosts has an average 
value of $\sim$4. This is in agreement with the redshift
dependent value of 3$-$7 that \citet{cha01} find for the ratio between the global comoving
dust-obscured
star-formation rate and the unobscured star-formation rate derived from the UV continuum.

\subsection{Mass and Age Estimates}

In this section, we 
calculate the SFR per unit stellar mass ($\dot{M}/M$) of the GRB host galaxies. Our K-band data
allows a better constraint on the stellar mass of the host galaxy than derived
by \citet{sok01} which was based in most cases on B, V, R, I photometry.
For this purpose, we used the newest version (year 2000) of the
population synthesis spectral energy distributions (SEDs) of
\citet{bc93} considering templates with both solar and 0.02 times solar metallicity.
A variety of star-formation histories of the form $\exp(-t/\tau)$ are selected
for the templates, ranging from
$\tau$ of 1 Myr which corresponds to a single, brief epoch of star-formation, to 
near-constant with a $\tau$ of 10 Gyr.
Screen extinction internal to the host galaxy was also incorporated
using the extinction curve of \citet{mat90} and the starburst extinction curve of \citet{cal00}.
The fits were performed by minimizing the sum of absolute errors weighted by the photometric
uncertainty at each of the wavelengths.
The parameters that are derived from the fits to the
observed magnitudes are the total stellar mass (M$_{\rm gal}$), internal
extinction (A$_{\rm V}$), age of the starburst ($t$),
template metallicity and {\it e}-folding timescale of the starburst ($\tau$). Of these, the metalli
city
and {\it e}-folding timescale are relatively unconstrained by the quality of the available photomet
ry.
The range of acceptable values for the three other parameters i.e. mass, age and extinction, are
obtained from fits
with different constraints e.g. with and without extinction, Galactic extinction and starburst
extinction, low metallicity and high metallicity.
The true statistical uncertainty in the parameters derived from
these fits is much larger as has been illustrated in
\citet{pap01}.

GRB970228: 
The host of GRB970228 has been found to have a redshift of 0.695 \citep{blo01b}.
It has a roughly circular/irregular morphology in HST images but displays
no evidence of an ongoing tidal interaction. 
The intrinsic color of the host is quite blue, rather typical of field galaxies and
suggestive of active star formation \citep{fru99}. 
However, analysis of the visible light spectrum by \cite{blo01} seems to suggest 
uniform star-formation if internal extinction is negligible, rather than a few
hundred Myr old starburst derived in the analysis of \citet{cas99}.
The derived luminosity at visible wavelengths is \ab0.1~L$_{*}$ i.e. quite sub-luminous.
Estimates of the SFR from the \ion{O}{2} line flux and UV luminosity
provide a value \ab0.5~M$_{\sun}$/yr \citep{blo01b}.
As a result, it has been classified as a late-type dwarf by \citet{blo01b} and the observed magnitudes
fit by a reddened Sc galaxy template by \citet{gal00}.
Estimates of the mass of the
galaxy have thus far been poorly constrained. 
If we assume no extinction or the starburst extinction curve, 
our template fits yield a mass for the host galaxy of
1.2$-$2$\times$10$^{8}~$M$_{\sun}$, $t\sim$ 40$-$80 Myr and $\tau=$ 1$-$20 Myr 
with A$_{\rm V}<0.2$~mag. 
If we instead adopt the
Galactic extinction curve, the derived
range of values depending on the metallicity are A$_{\rm V}\sim$0.2$-$0.9~mag,
M$_{\rm gal}\sim$1.6$-$2.5$\times10^{8}$~M$_{\sun}$, $t=$ 20$-$140 Myr
and $\tau=$ 20$-$70 Myr. 
This latter estimate of A$_{\rm V}$ agrees with that derived from the
$\beta$-slope technique. However, for this galaxy, 
both the B-band and V-band photometry used to
derive the UV-slope have large uncertainties associated with them. So the 
amount of dust extinction in this galaxy
derived from the $\beta$-slope technique is somewhat uncertain.
Irrespective of this, the $\dot{M}/M$ ratio for the host is rather high while the
age of the stellar population in the best-fitting templates are low, suggestive
of a recent starburst in this dwarf galaxy (Figure 2).
For reference, the brightness of the galaxy 2.5$\arcsec$ to the NE of the host is 
K$=21.6\pm0.1$~mag.

GRB970508: 
\citet{blo98} confirmed that the redshift of the GRB970508 host galaxy is 0.835.
HST/STIS observations \citep{fru00} revealed that
the galaxy is quite compact and 
that the GRB was centered within $\sim$70~pc of the nucleus of the galaxy.
The host has an intrinsic luminosity at visible wavelengths of \ab0.1~L$_{*}$,
again suggesting a dwarf galaxy with a UV-derived SFR of 0.2$-$1.4~M$_{\sun}$/yr.
In the absence of extinction, we find the mass of the galaxy to 
be $3.4\times10^{8}~$M$_{\sun}$ with
$t\sim$ 100~Myr and $\tau=$1$-$10 Myr.
If we consider the Galactic extinction curve, we find that the observed 
flux densities are best fit by a low
metallicity template with
A$_{\rm V}\sim0.8$~mag, M$_{\rm gal}\sim8\times10^{7}~$M$_{\sun}$,
$t=$ 7 Myr and $\tau$=1 Myr. This value is in good agreement with the estimate 
of internal extinction
derived from the $\beta-$slope technique.

There are three other possibly unrelated galaxies in the vicinity of the host
which are denoted by G1, G2 and G3 in \citet{zha98}. For reference, we derive the brightness
of G1, G2 and G3 to be
K=22.1$\pm$0.1~mag, 20.5$\pm$0.1~mag and 21.6$\pm$0.1~mag respectively. 
There is a fourth object $\sim$6$\arcsec$
N-NW of the host with K=21.3$\pm$0.1~mag which we call G4. 
To the best of our knowledge, the redshifts of these individual objects have not been measured. 

GRB971214: HST imaging revealed a galaxy of irregular morphology with
rest frame luminosity L$_{\rm UV}$\ab0.2~L* \citep{ode98}.
It has a  measured redshift of 3.418 \citep{kul98}.
The results from our template fitting method are quite uncertain since there are measurements at only
3 bands of which the photometry in the V-band differs by more than 1 magnitude between \citet{sok01} 
and \citet{ode98}. 
We adopt a weighted geometric mean of the two V-band measurements which results
in V$=$26.2$\pm$0.3~mag.
In addition, interstellar absorption features, Ly$\alpha$ emission and 
the Lyman forest could be responsible for contaminating the V and R-band photometry for which
we have applied an empirical correction in the derivation of the UV-slope but not
to the individual photometry values.
Since the age of the Universe at $z=3.418$ is 1.7~Gyr, this places
an additional constraint on the template ages in our fits to the multiband photometry.
Fits without extinction result in M$_{\rm gal}\sim8\times10^{9}~$M$_{\sun}$
with $t\sim$ 200~Myr and $\tau\sim$30~Myr. Fits including extinction (A$_{\rm V}\sim$0.5~mag)
suggest higher masses
and older stellar populations with M$_{\rm gal}\sim$1$-$5$\times10^{10}~$M$_{\sun}$, 
$t\sim$100$-$700~Myr, $\tau\sim0.03-2$~Gyr. 
The UV slope suggests
an average A$_{\rm V}\sim1.8$~mag for the UV light, an infrared 
luminosity of $2\times10^{12}~{\rm L}_{\sun}$
and a SFR of $\sim$350~M$_{\sun}$/yr.
This is quite similar to the
3$\sigma$ upper limit of 3~mJy obtained for this galaxy at 850~$\mu$m which at $z=3.418$,
translates to L$_{\rm IR}$ for the galaxy of 3.3$\times10^{12}~$L$_{\sun}$ \citep{Smi99}.
However, the best-fitting exponentially decaying
star-formation history results in a galaxy mass that exceeds the derived stellar mass. 
While it is possible that this is a massive galaxy undergoing a violent starburst,
we conclude that better multiband photometry at visible wavelengths is 
required to determine the true parameters of this galaxy.
For reference, the brightness of the galaxy located 2$\arcsec$ to the W-NW is K$=21.5\pm0.1$~mag.
 
GRB980326: The HST/STIS observations of \citet{fru01a} yielded a tentative 
detection of the host galaxy with V=$29.3\pm0.3$~mag
but the object is too faint for any other characteristics to be derived.
When the HST image is aligned with ours, the flux in our image at the position of the
galaxy corresponds to K$=22.9\pm0.4$~mag. The presence of bright stars in the frame and high
residuals in that part of the image reduces the credibility of this detection. 
For reference, the object 2.4$\arcsec$ to the East has K$=21.1\pm0.1$~mag.

GRB980329: \citet{hol00b} provided tentative evidence for a host galaxy located
about 0.5$\arcsec$ southwest of the position of the optical transient with
R$\approx28$~mag. Again, the object is too faint at visible wavelengths
and undetected in the near-infrared for any characteristics of the host
to be derived. However, it does seem that the \citet{sgd98b} measurement of the GRB980329
host is contaminated by the optical transient.

GRB980519:
Visible light observations of GRB980519 revealed a faint underlying galaxy as well as a companion 
galaxy 1.5$\arcsec$ to the SW \citep{hol00a}. 
The host has V=$28\pm0.3$~mag while the companion has V=$27\pm0.1$~mag. 
Our infrared observations confirm 
the detection of the combined system with K$=22.5\pm0.3$~mag. 
However, the object is at the limit of detectability and as a result we are unable to
derive any physical parameters for the host.

GRB980613: 
The GRB apparently originated in an interacting system at $z$=1.097, where the host has
at least 2 faint galaxy companions and 2 
bright ones all of which are seen in the visible image and denoted A (host), B, C, D, E \citep{sgd00}. 
We found values of K=$21.7\pm0.2$~mag, 21.6$\pm$0.2~mag, 20.2$\pm$0.2~mag, 20.3$\pm$0.2~mag
and 22.3$\pm$0.2~mag for A, B, C, D and E respectively, in reasonable agreement with the 
K-band values
of \citet{sgd00}. Components A and E are relatively blue compared to components B, C and D.
The SFR as derived from the visible/UV emission is $\sim$5 M$_{\sun}$/yr and 
it's luminosity is lower than present-day L$_{*}$ galaxies leading \citet{sgd00} to conclude
that the galaxy is undergoing a mild starburst.
Fits to the multiband photometry
without extinction are quite poor. Fits including extinction
(A$_{\rm V}=1-2$~mag) yield a mass of M$_{\rm gal}=0.5-2.6\times10^{9}$~M$_{\sun}$, $t=$ 3$-$8~Myr and
$\tau$=1$-$30~Myr. The lower limit to the obscured SFR derived from the $\beta-$slope
technique is 70~M$_{\sun}$/yr with A$_{\rm V}=1.8$~mag. Alternatively,
applying an extinction-correction to the SFR derived from the UV continuum results in
60~M$_{\sun}$/yr, in excellent
agreement with our $\beta$-slope value.

GRB980703: 
This is one of the brightest hosts seen in our sample and is
located at $z=0.966$ \citep{sgd98a}. HST/STIS imaging has determined that it is a very compact galaxy
with a faint companion galaxy $\sim$2$\arcsec$ to the S-SE \citep{hol01}.
If the internal extinction in the host is assumed to be negligible,
the SFR derived from the UV continuum and the OII line
flux is in the range 8-20 M$_{\sun}$/yr \citep{sgd98a}. 
While there is evidence for extinction in this galaxy based on the visible light spectrum of the
host and analysis of the X-ray/visible/near-infrared light curve of the transient \citep{cast99},
the range of values span a wide range, from A$_{\rm V}=0.3$~mag to 2.2~mag. The galaxy has also been detected
at radio wavelengths by \citet{berg01}. The radio luminosity of the galaxy yields a SFR
from massive stars (M$>5$~M$_{\sun}$) of $\approx$90~M$_{\sun}$/yr. This 
has been extrapolated using a Salpeter mass function 
to yield a total SFR of 500~M$_{\sun}$/yr. The derived 
far-infrared luminosity, adopting the relation derived by \citet{Con92}, is 
greater than 10$^{12}$~L$_{\sun}$, clear 
evidence that this galaxy is a ULIG. Furthermore, the position of the GRB is very close to the 
nucleus of the galaxy, suggestive of an origin in a nuclear starburst.

Template fits to the multiband photometry with no extinction are rather poor.
If we include extinction, A$_{\rm V}$ spans the range 0.3$-$1.2~mag, $t$=10$-$30~Myr,
M$_{\rm gal}$=1$-$4.6$\times10^{9}$~M$_{\sun}$ and $\tau$=1$-$20~Myr.
The $\beta-$slope technique yields a lower limit of A$_{\rm V}\sim$0.9~mag and an obscured SFR
of 30 M$_{\sun}$/yr. Assuming an  A$_{\rm V}\sim1$~mag results in an extinction corrected SFR
from the UV continuum of 45 M$_{\sun}$/yr. The SFR estimate from the extinction-corrected
UV continuum is in excellent
agreement with that obtained from the sum of 
the $\beta-$slope technique and the observed UV continuum.
Furthermore, this value is within a factor of 2 of the SFR in massive stars derived from the radio
luminosity. This curious agreement between the SFR in massive stars
and our extinction-corrected values provides weak evidence for the star-formation in this galaxy to
be biased towards the high mass end. Alternatively, it is possible that much of the
star-formation in this galaxy is in optically thick regions and as a result, insensitive
to measurements in the UV.

GRB981220:
A radio transient presumably associated with GRB981220
was detected within a few days of the burst \citep{gal98}.
\citet{blo99c} found a variable source in their visible light images
located at the position of the radio transient
and which was therefore presumed to be the visible light afterglow superposed on the host.
Later VLBA observations of the radio transient
revealed that it has a core-jet
morphology extending to the S-W \citep{tay99}.
This was therefore interpreted to be an intraday variable source unrelated to the GRB.
However, the brightness of this source has not varied over the two
epochs of the infrared observations and it is extended
in the NE-SW direction in seeing conditions of 0.5$\arcsec$.
There is also an excess of flux 1.3$\arcsec$ to the west which
we refer to as a ``companion".
The brightness of the object and companion are
K=19.0$\pm$0.1~mag and 22.0$\pm$0.3~mag respectively.
Comparison with published visible light photometry \citep{blo99c} indicates
the object is very red (R$-$K=7.4$\pm$0.5~mag) compared to the other
host galaxies which would indicate it to be either a unique host
or a source that has varied between the time of the visible light and infrared
observations. The companion (object 'K' in Bloom et al.)
is relatively bluer at (R$-$K=3.5$\pm$0.5~mag)
and similar in color to the other hosts.
While it is possible
that this is an interacting system, the association
of these objects to GRB981220 is unclear but probably spurious since the discussed source
is outside the refined IPN error box of the burst. 

GRB990123: 
HST/STIS imaging of the host galaxy of GRB990123 indicates that
it has a disturbed morphology \citep{blo99a, fru99b}
with knots of star-formation which have been denoted as A, A1, A2, \#1 and B. 
Analysis of the STIS data by \citet{hh99} provided
magnitudes of V=28.1$\pm$0.3~mag for knots A1, A2 and \#1. The host which is presumed to be knot A
since it is the brightest object in the system, is located at a redshift of 1.6 \citep{and99, kul99} and has
a V magnitude of 24.25$\pm$0.2~mag \citep{fru99b, hh99}. 
The derived SFR for knot A is \ab4~M$_{\sun}$/yr and its blue luminosity L$_{\rm B}$\ab0.5L*.
In comparison, the knots are thought to have a SFR of only about 0.1-0.2~M$_{\sun}$/yr \citep{hh99}.
The proximity of the different knots makes it difficult to resolve them and perform
photometry on each component in the K-band data. So, we provide an integrated
brightness of K=21.7$\pm$0.2 mag for the entire system. 
For reference, field 
objects L and M \citep[see][]{blo99a} have K=19.56$\pm$0.1~mag and 19.44$\pm$0.1 mag
respectively.

It has also been shown that the galaxy is quite blue but not very luminous
for galaxies at that redshift. Interestingly, our estimate of the internal
extinction in this host is quite small.
We find A$_{\rm V}$=0.5~mag based on the $\beta$-slope technique while the template
fits to the multiband photometry yield A$_{\rm V}$=0~mag, M$_{\rm gal}$=$3.6\times10^{9}$~M$_{\sun}$
with $t=$ 140$-$180 Myr and $\tau\sim$ 30 Myr. 
This could be interpreted as evidence that not all interacting
systems necessarily result in strong star-formation in dust-enshrouded regions. On the other hand,
it is equally likely that there are regions of strongly obscured star-formation that are
located in the midst of the detected knots of emission but they are optically thick
to UV light. 
Good observational evidence for such a hypothesis comes from the ``Antennae'' galaxy
which shows strong mid-infrared emission arising from warm dust in regions that
are inconspicuous in UV light \citep{Mir98}.
So it is possible that the integrated SFR for the system is actually
much higher than the sum of it's parts.
It should also be noted that for this galaxy, much of the early photometry for the host galaxy
was performed by masking or fitting the point source corresponding to the transient. This 
seems to induce a significant inaccuracy in the early-time photometry values as is illustrated
in the difference between the values of \citet{sok01} and \citet{fru99b} and between our near-infrared
values and those derived by \citet{blo99a}.

GRB991208: 
The near-infrared image of GRB991208 reveals the host and a companion galaxy `A'
about 1$\arcsec$ SE of the host which is also seen in the HST/STIS
image \citep{cast01}.
There appears to be an additional 
object `B' that is 2.5$\arcsec$ East and slightly North of the host
which is not visible in the STIS image
but it is not clear from the image if these objects are connected
by a tidal stream. The brightness of the host galaxy, galaxy `A' and `B' are
K=21.7$\pm$0.2~mag, 23.1$\pm$0.4 mag and 22.4$\pm$0.2~mag respectively.
The V$-$K color of the host, which is compact in the HST images, is 3.0$\pm$0.3~mag 
\citep{cast01}. If all these galaxies are at a $z\sim0.7$ as 
inferred for the GRB \citep{dod99, sgd99}, their projected physical separation is 
about 7~kpc/arcsec. It is necessary
to measure the redshift of these three objects to establish any dynamical interaction
between them. \citet{cast01} inferred from the broadband photometry that the galaxy is
not exceptionally bright and has a SFR of \ab5$-$18~M$_{\sun}$/yr.
The best fits to the multiband photometry result from low metallicity
templates with little or no extinction. 
The derived mass of the host is 8.6$\times10^{8}$~${\rm M}_{\sun}$ for $\tau\approx$70~Myr
and $t\sim$300~Myr. The low value of extinction is inconsistent with our estimate from
the $\beta$-slope technique. However, we conclude that the UV-slope result for this galaxy
is uncertain
for reasons mentioned in Section 3.2.

GRB000301C: The HST/STIS image of \citet{fru01} with the transient was aligned
with our image and photometry performed at the position of 
the transient. This yielded a 
magnitude of K=23.0$\pm$0.5~mag. We do not have full confidence in this detection since the
observed brightness of the host 
is significant only at the \ab2.5$\sigma$ level. However, if the measurement is real,
it must be the host galaxy because the transient should have faded to a level
fainter than \ab24~mag \citep{rho01}. When compared
with the V-band photometry of Fruchter \& Vreeswijk (2001), the host galaxy has 
an observed V$-$K color of 5.0$\pm$0.6~mag which is redder than most hosts.
The companion galaxy located at an angular separation of 2.3$\arcsec$ to the north-west
(18 kpc projected distance)
has a brightness of K=$19.8\pm0.1$~mag which results in R-K$=4.5\pm0.3$~mag. 
However, there is no evidence of any tidal debris between the burst position and this galaxy.
For reference, the bright star located to the south-west has a brightness of K=16.02$\pm$0.05~mag.

Although the masses derived for individual GRB host galaxies from the different fits
span a relatively narrow range,
the statistical uncertainty in the derived masses can be quite large, about an order of magnitude
depending on the photometric uncertainties. In addition, fitting a single template to the multiband
photometry often results in the lowest mass system because of the low mass-to-light ratio
of a young stellar population. \citet{pap01} have demonstrated that including an older
stellar component in the fits typically results in an upper limit on the mass which is a
factor of 2$-$3 higher than that derived from fitting a single template. Irrespective of 
this, the sense of Figure 2 remains the same even if the masses of the GRB hosts were corrected
upwards by a factor of 3.

In addition, fits which include stellar populations of two different ages
result in a lower age for the component of stars formed in the more recent starburst.
The presence of young stellar populations ($<$10$^{8}$~yr) in GRB host galaxies,
combined with their high star-formation rate per unit stellar mass seems to suggest
that most GRB host galaxies have undergone a recent starburst. The age of the stellar populations
is comparable to the lifetimes of massive ($>$10~M$_{\sun}$) stars providing some evidence
for GRBs originating from collapsars rather than from the inspiral and merger of
double degenerate objects which can typically take a few hundred Myr from the onset 
of star-formation \citep{fry99}.

\section{Conclusions}

Of the 12 putative gamma-ray burst positions that were observed, all except
GRB981220 had definite identification of a visible light transient associated with the burst.
For the remaining 11 bursts,
visible light observations of the transient position detected
an underlying host galaxy after the transient had faded.
Our observations reveal that
at least 6 of the 11 hosts appear to be in systems with another galaxy within
a projected angular separation of 2.5 $\arcsec$.
We have derived extinction corrected star-formation rates for 7 of the hosts which
have measured redshifts using the UV-slope technique. In addition, we have derived
other physical parameters of the host galaxies such as their mass, age of the starburst,
and internal extinction based on fits to the multiband photometry between visible
and near-infrared wavelengths.
We find that the extinction corrected star-formation rates are significantly higher
than the estimates derived from rest-frame UV continuum emission and the \ion{O}{2}
line strengths. More interestingly, the star formation rates per unit stellar mass ($\dot{M}/M$) of
these galaxies is higher than for typical nearby starburst galaxies.
In addition, the template fits to 4 of the objects provide
evidence of a young stellar population of age about 10$-$50~Myr, a typical
timescale for the formation of a `collapsar' through the core collapse of an isolated
massive star and explosion of the resultant black hole/accretion disk system.
The high incidence of GRB hosts in close pairs of galaxies and high $\dot{M}/M$ values
strengthen the argument that the progenitors of at least some of the long duration GRBs are
high-mass stars in starbursts. High quality multiband photometry of
a statistically large sample of host galaxies will be
required to assess if the stellar mass function in high redshift starbursts is biased
towards the high mass end.

\acknowledgements

We are very grateful to Gerry Neugebauer, David Hogg and Mark Morris for their 
support of this project.
We thank Casey Papovich for his assistance with the Bruzual \& Charlot
template spectral energy distributions. 
We also wish to acknowledge Scott Barthelmy for operating the GCN System
which has been a useful resource for the entire GRB community.
The comments of an anonymous referee are also much appreciated.
This work is partly funded by NASA grant NAG5-3042 and is based on 
observations made at the W. M. Keck 
Observatory which is operated as a scientific partnership between the
University of California, the California Institute of Technology and the
National Aeronautics and Space Administration.

\begin{figure}
\begin{minipage}{1.8in}
\epsfxsize=1.8in
\epsfbox{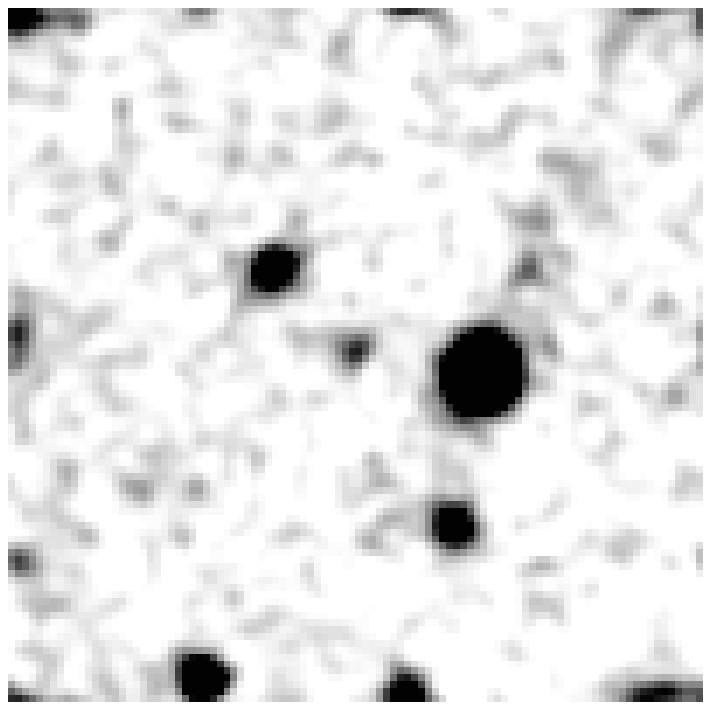}
\end{minipage}
\begin{minipage}{1.8in}
\epsfxsize=1.8in
\epsfbox{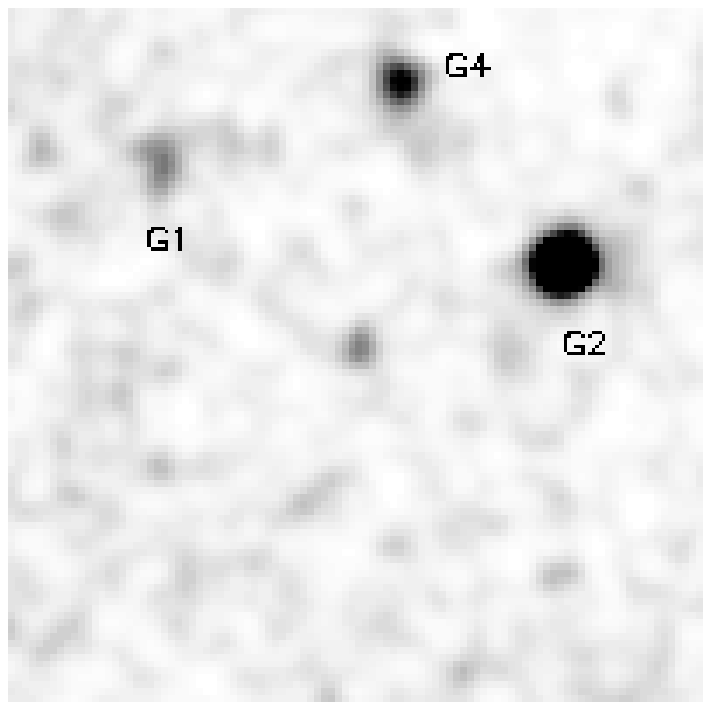}
\end{minipage}
\begin{minipage}{1.8in}
\epsfxsize=1.8in
\epsfbox{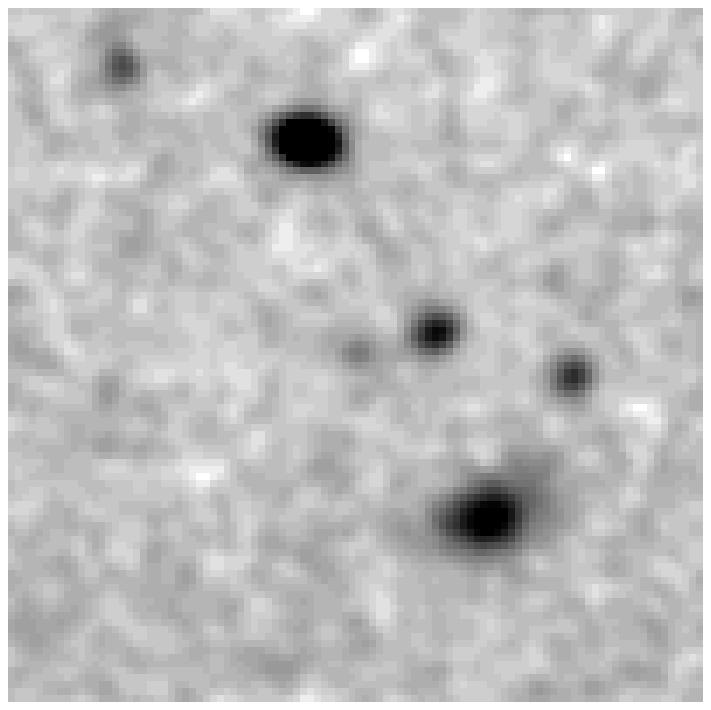}
\end{minipage}

\begin{minipage}{1.8in}
\epsfxsize=1.8in
\epsfbox{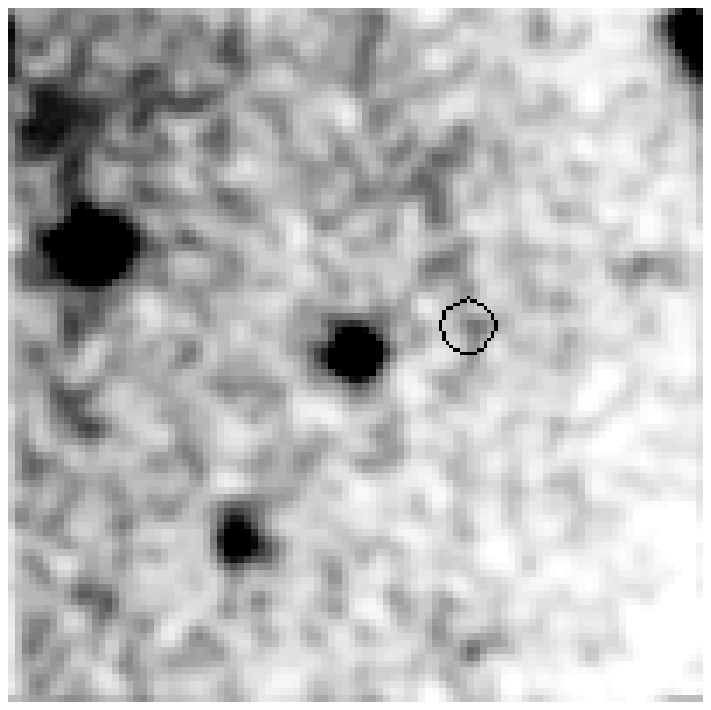}
\end{minipage}
\begin{minipage}{1.8in}
\epsfxsize=1.8in
\epsfbox{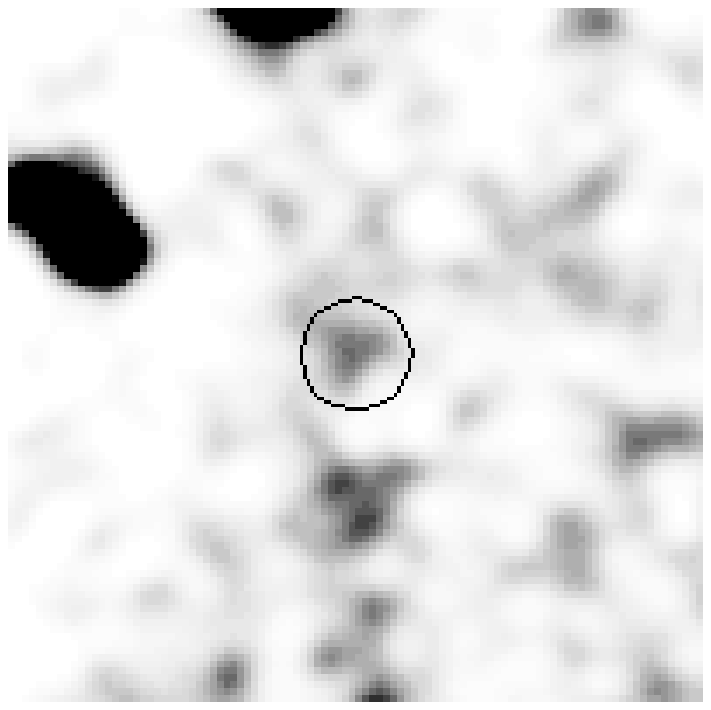}
\end{minipage}
\begin{minipage}{1.8in}
\epsfxsize=1.8in
\epsfbox{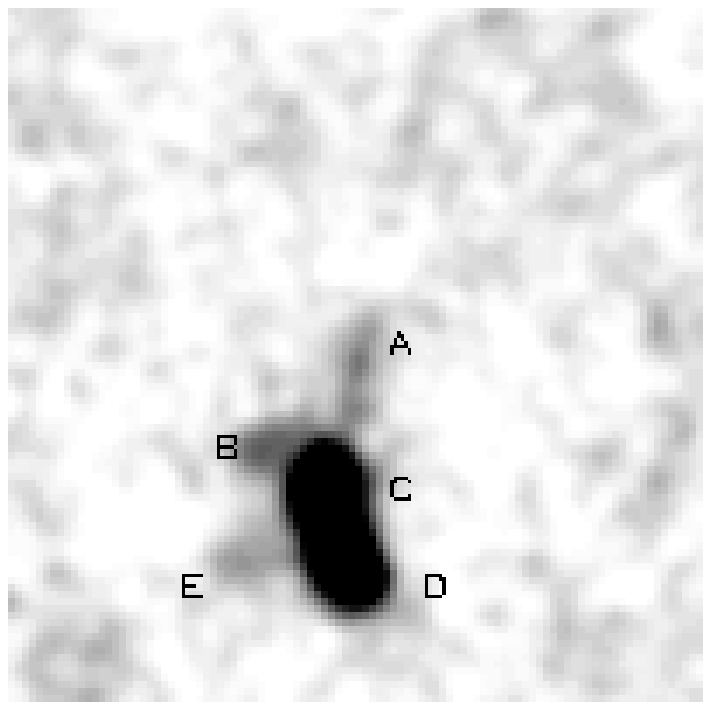}
\end{minipage}

\begin{minipage}{1.8in}
\epsfxsize=1.8in
\epsfbox{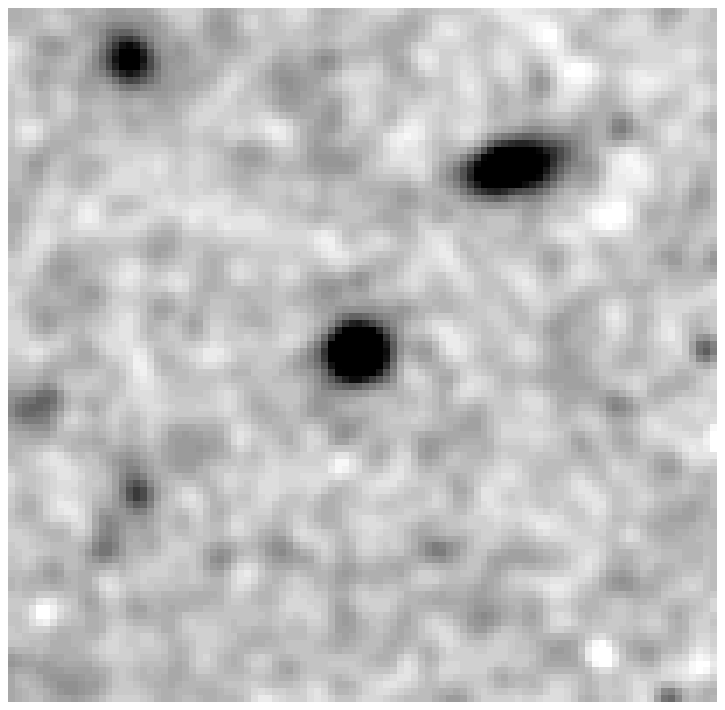}
\end{minipage}
\begin{minipage}{1.8in}
\epsfxsize=1.8in
\epsfbox{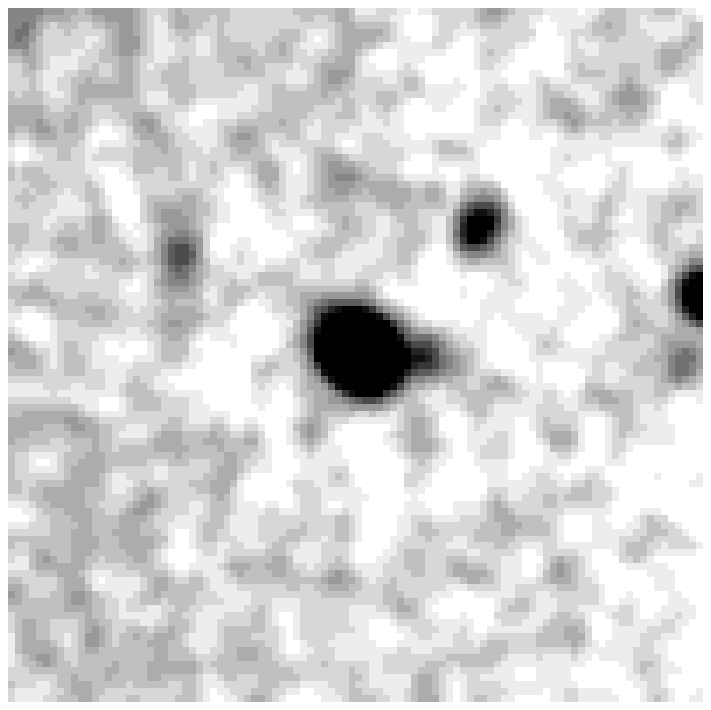}
\end{minipage}
\begin{minipage}{1.8in}
\epsfxsize=1.8in
\epsfbox{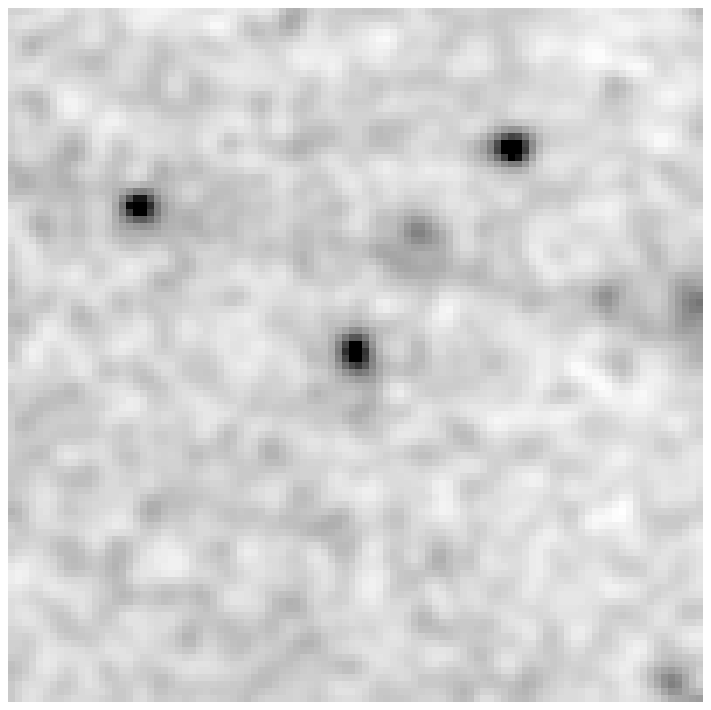}
\end{minipage}

\begin{minipage}{1.8in}
\epsfxsize=1.8in
\epsfbox{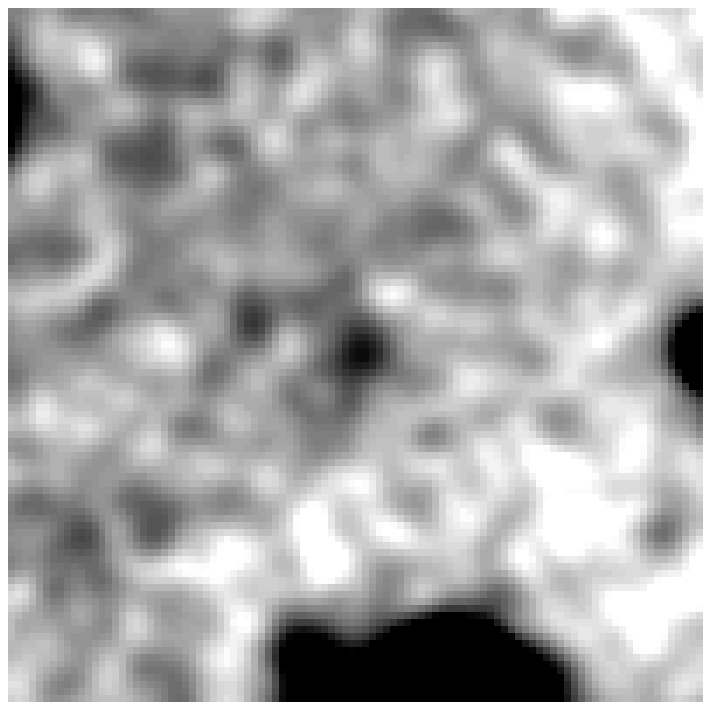}
\end{minipage}

\caption{
Deep K-band images of selected GRB host galaxies where North is up and
East to the left. The images are
15$\arcsec$ on a side and the host is either located at the center of the field or
marked with a circle.
In many cases a companion galaxy can be seen within about
2.5$\arcsec$ radius of the host. From left to right, top to bottom,
the images are of 970228, 970508, 971214, 980326,
980519, 980613, 980703, 981220, 990123, 991208.}
\end{figure}

\begin{figure}[ptb]
\plotone{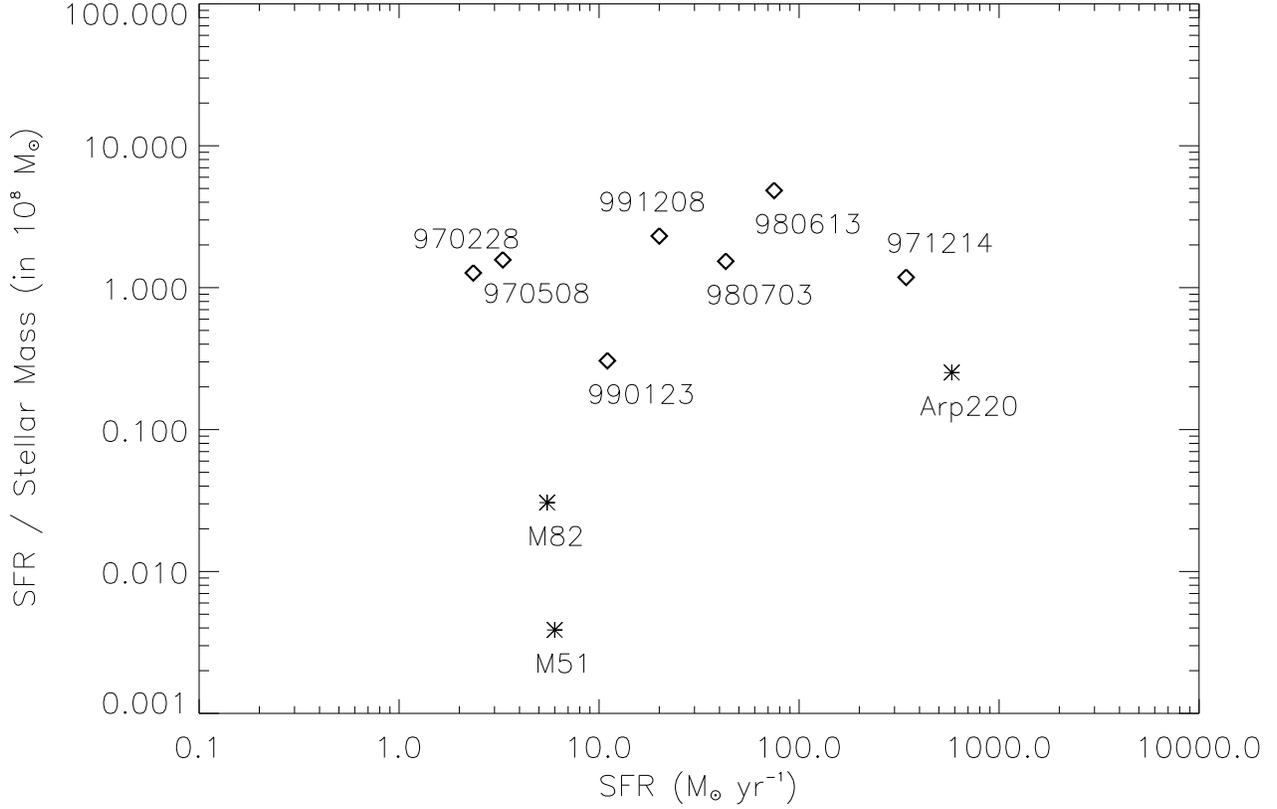}
\caption{Total (obscured+unobscured) star formation rates for the gamma-ray burst host galaxies
derived using the $\beta-$slope technique
plotted against the ratio between the star formation rates and the midpoint of the range
of stellar masses derived
from fits to the multiband photometry
(Table 3 has the stellar mass from the best-fitting templates).
The uncertainty on the GRB971214 point is quite large because of the lack of good photometry
at multiple wavelengths.  
The star-formation rates for the GRB hosts are lower 
limits (see text), so the data points are likely to move higher and to the right.
Estimates of the stellar mass typically have 95\% confidence intervals that span an order of magnitude.
Also plotted are the corresponding values
for two prototypical starbursts Arp220 and M82 and the relatively quiescent Sbc galaxy M51 
\citep{Sil98}.
GRB hosts have star-formation rates per unit stellar mass much higher than local starbursts.
}
\end{figure}

\begin{deluxetable}{ccccccc}
\singlespace
\tablecaption{Summary of observations}
\tablehead{
\colhead{Source} &
\colhead{RA} &
\colhead{Dec} &
\colhead{UT Date\tablenotemark{d}} &
\colhead{On source time} &
\colhead{Seeing} &
\colhead{Sensitivity\tablenotemark{c}} \\
\cline {2-3}
\colhead{} & \multicolumn{2}{c}{(J2000)} & \colhead{} & \colhead{seconds} &
\colhead{FWHM} & \colhead{3$\sigma$}
}
\tablewidth{0pt}
\tablecolumns{7}
\startdata
GRB970228 & 05:01:46.7 & 11:46:53 & Jan 11.4, 12.3 & 12840 & 0.51$\arcsec$ & 23.7 \\
GRB970508 & 06:53:49.5 & 79:16:20 & Nov 30.5 & 7080\tablenotemark{a,b} & 0.47$\arcsec$ & 22.6 \\
          &             &         & Jan 12.5 & 7200 & 0.6$\arcsec$ & 23.2 \\
GRB971214 & 11:56:26.4 & 65:12:01 & Jan 12.6 & 7200 & 0.42$\arcsec$ & 23.4 \\
GRB980326 & 08:36:34.3 & -18:51:24 & Jan 11.5, 12.4 & 11625 & 0.63$\arcsec$ & 23.0 \\
GRB980329 & 07:02:38.0 & 38:50:44 & Dec 2.4 & 5220\tablenotemark{a,b} & 0.4$\arcsec$ & 22.8 \\
GRB980519 & 23:22:21.5 & 77:15:43 & Jan 11.2, 12.2 & 10200 & 0.9$\arcsec$ & 23.0 \\
GRB980613 & 10:17:57.6 & 71:27:26 & Jan 11.6 & 7275 & 1.0$\arcsec$ & 23.1 \\
GRB980703 & 23:59:06.7 & 08:35:07 & Jan 11.2 & 1350 & 0.55$\arcsec$ & 22.4 \\
GRB981220 & 03:42:28.9 & 17:09:15 & Jan 11.3 & 2700 & 0.5$\arcsec$ & 22.8 \\
	  &	       & 	  & Jan 30.3 & 2400 & 0.84$\arcsec$ & 22.5 \\
GRB990123 & 15:25:30.3 & 44:45:59 & Jan 29.7 & 1800 & 0.8$\arcsec$ & 23.0 \\
 & & & Apr 28.6 & 2340 & 0.75$\arcsec$ & 22.3 \\
 & & & Apr 29.6 & 5700\tablenotemark{b} & 0.5$\arcsec$ & 23.0 \\
GRB991208 & 16:33:53.5 & 46:27:21 & Apr 19.5 & 2880 & 0.8$\arcsec$ & 22.5\\
GRB000301C & 16:20:18.6 & 29:26:36 & Apr 19.6 & 3600 & 0.7$\arcsec$ & 22.6\\
\enddata
\tablenotetext{a}{Observations made in K$_{\rm s}$~filter}
\tablenotetext{b}{Non-photometric conditions}
\tablenotetext{c}{In a beam of diameter 1.5$\arcsec$. The sensitivity was calculated from the
standard deviation in the background values.}
\tablenotetext{d}{Dates are Nov 29-Dec 1 1998, Jan 11-12 1999,
Jan 29-30 1999, Apr 28-29 1999 and Apr 19 2000}
\end{deluxetable}

\begin{landscape}
\begin{deluxetable}{cccccccccc}
\tablecaption{Observed Parameters of GRB Host Galaxies}
\tablehead{
\colhead{GRB} &
\colhead{Redshift} &
\multicolumn{6}{c}{Photometry} & 
\colhead{Angular Size} & \colhead{Reference}\\
\cline{3-8}
\colhead{} & \colhead{} & \colhead{B} & \colhead{V} & \colhead{R} &
\colhead{I} & \colhead{K\tablenotemark{4}} & \colhead{Others} & \colhead{} 
}
\tablewidth{0pt}
\scriptsize
\tablecolumns{9}
\startdata
970228 & 0.695 & 26.4$\pm$0.3 & 25.8$\pm$0.3 & 25.2$\pm$0.3 & 24.7$\pm$0.2 & 22.6$\pm$0.2 & H$=$23.3$\pm$0.1 & 0.8$\arcsec$ & 1, 2, 3\\ 
970508 & 0.835 & 25.9$\pm$0.2 & 25.3$\pm$0.2 & 25.1$\pm$0.2 & 24.1$\pm$0.3 & 22.7$\pm$0.2 & & 0.6$\arcsec$ & 4, 5 \\
971214 & 3.418 & $>$26.8 & 26.6$\pm$0.2 & 25.6$\pm$0.2 & $>$24.5 & 22.4$\pm$0.2 & & 0.4$\arcsec$ & 6, 7\\
980326 & ...   &  &  & 29.3$\pm$0.3 & & 22.9$\pm$0.4\tablenotemark{1} &  & & 8\\
980329 & ...   &  &  & 28.0$\pm$0.3 & & $>$22.8 & & & 9\\
980519 & ...   &  &  & 28.0$\pm$0.3 & $>$24.5\tablenotemark{2} & 22.5$\pm$0.3\tablenotemark{1} & & 0.5$\arcsec$ & 10, 11\\
980613 & 1.097 & 25.1$\pm$0.3 & 24.2$\pm$0.2 & 23.8$\pm$0.2 & 23.4$\pm$0.1 & 21.7$\pm$0.1 & & 2$\arcsec$ & 5, 12 \\
980703 & 0.966 & 23.4$\pm$0.2 & 23.0$\pm$0.1 & 22.6$\pm$0.1 & 22.3$\pm$0.2 & 19.6$\pm$0.1 & J=21.1$\pm$0.2 & $<$0.6$\arcsec$ & 5, 13, 24, 25\\
& & & & & & & H=20.7$\pm$0.3 & \\
981220 & ...   & $>24.6$ & $>24.2$ & 26.4$\pm$0.5\tablenotemark{3} & $> 23.0$ & 19.0$\pm$0.1 & & $<$ 0.7$\arcsec$ & 14, 15, 16\\
990123 & 1.6 & 25.0$\pm$0.2 & 24.6$\pm$0.2 & 24.5$\pm$0.1 & 24.1$\pm$0.3 & 21.7$\pm$0.3 & & 1$\arcsec$ & 5, 17, 18, 19 \\
991208 & 0.706 & 25.2$\pm$0.2 & 24.6$\pm$0.2 & 24.3$\pm$0.2 & 23.3$\pm$0.2 & 21.7$\pm$0.2 & & $<0.1\arcsec$ & 5, 20, 21\\
000301C & 2.034 & & & 28.0$\pm$0.3\tablenotemark{1} & & 23.0$\pm$0.5\tablenotemark{1} & & & 22, 23 \\
\enddata
\tablenotetext{1}{$\lesssim$3$\sigma$ detection}
\tablenotetext{2}{Gunn band}
\tablenotetext{3}{Possible afterglow contamination}
\tablenotetext{4}{This paper}
\end{deluxetable}
\end{landscape}

\begin{landscape}
\begin{deluxetable}{cccclcccc}
\tablecaption{Derived Parameters of GRB Host Galaxies}
\tablehead{
\colhead{Source} &
\colhead{Galactic Extinction\tablenotemark{a}} & ${\rm L_{UV-NIR}}$ & ${\rm L_{IR}}$ & Galaxy Type & UV SFR & 
$\beta$-slope SFR & M$_{\rm gal}$ & $\dot{M}/M$ \\
\colhead{} & \colhead{E(B-V)} & 10$^{9}$ L$_{\sun}$ & 10$^{9}$ L$_{\sun}$ & & M$_{\sun}$/yr & M$_{\sun}$/yr &
10$^{8}$~M$_{\sun}$ & 10$^{-8}$~yr$^{-1}$ \\ 
}
\tablewidth{0pt}
\tablecolumns{7}
\startdata
GRB970228 & 0.20 & 0.9 & 9.3 & Disk ? & 0.5$-$1 & 1.6 & 1.6 & 1.5 \\
GRB970508 & 0.04 & 1.2 & 14 & Elliptical & 0.2$-$1.4 & 2.5 & 0.8 & 4.1 \\
GRB971214 & 0.02 & 25 & 2000 & Irregular & 1$-$5 & 340 & 300 & 1.1 \\
GRB980326 & 0.08 & ... & ... & ... & ... & ... & ... & ... \\
GRB980329 & 0.15 & ... &... & ... & ... & ... & ... & ... \\
GRB980519 & 0.35 & ... &... & Interacting ? & ... & ... & ... & ... \\
GRB980613 & 0.09 & 4 & 400 & Interacting & 5 & 70 & 10 & 7.5 \\
GRB980703 & 0.06 & 18 & 170 & Unresolved & 6-20& 30 & 35 & 1.2 \\ 
GRB981220 & 0.20 & ... & ... & AGN ? & ... & ... & ... & ... \\
GRB990123 & 0.03 & 13 & 38 & Irregular & 4 & 7 & 36 & 0.3 \\
GRB991208 & 0.02 & 0.9 & 45 & Interacting ? & 5-18 & $<$8 & 9 & 2.1 \\
GRB000301C & 0.05 & 3.9 & ...  &   ...         &   ...   & ... & ... & ... \\
\enddata
\tablenotetext{a}{From the dust maps of Schlegel, Finkbeiner \& Davis (1998).}
\end{deluxetable}
\end{landscape}

\end{document}